\documentclass[letterpaper, 10 pt, conference]{ieeeconf}  

\IEEEoverridecommandlockouts                              
\usepackage{cite}
\usepackage{amsmath,amssymb,amsfonts}
\usepackage{algorithm}
\usepackage{algorithmicx}
\usepackage{algpseudocode}
\usepackage{graphicx}
\usepackage{textcomp}
\usepackage{xcolor}
\usepackage{balance}

\def\BibTeX{{\rm B\kern-.05em{\sc i\kern-.025em b}\kern-.08em
    T\kern-.1667em\lower.7ex\hbox{E}\kern-.125emX}}

\newtheorem{definition}{Definition}
\newtheorem{lemma}{Lemma}
\newtheorem{proposition}{Proposition}

\title{\LARGE \bf
Uncertainty Removal in Verification of Nonlinear Systems against Signal Temporal Logic via Incremental Reachability Analysis
}

\usepackage{todonotes}

\author{Antoine Besset$^{*}$, Joris Tillet$^{*}$ and Julien Alexandre dit Sandretto$^{*}$
\thanks{$^{*}$U2IS, ENSTA, Institut Polytechnique de Paris, Palaiseau, France}}
        
\begin{document}

\maketitle
\thispagestyle{empty}
\pagestyle{empty}
\maketitle


\begin{abstract}
%
A framework is presented for the verification of Signal Temporal Logic (STL) specifications over continuous-time nonlinear systems under uncertainty. Based on reachability analysis, the proposed method addresses indeterminate satisfaction caused by over-approximated reachable sets or incomplete simulations. STL semantics is extended via Boolean interval arithmetic, enabling the decomposition of satisfaction signals into unitary components with traceable uncertainty markers. These are propagated through the satisfaction tree, supporting precise identification even in nested formulas.
To improve efficiency, only the reachable sets contributing to uncertainty are refined, identified through the associated markers. The framework allows online or offline monitoring to adapt to incremental system evolution while avoiding unnecessary recomputation. A case study on a nonlinear oscillator demonstrates a significant reduction in satisfaction ambiguity, highlighting the effectiveness of the approach.
\end{abstract}
\section{Introduction}
Ensuring the reliability and safety of dynamical and controlled systems is critical across domains such as autonomous systems, cyber-physical systems, and industrial automation. Despite careful design, uncertainties like model inaccuracies or external disturbances can cause deviations from expected behavior. Model-based verification methods support system analysis by simulating state evolution and identifying discrepancies between intended and actual behavior~\cite{dvorak_model-based_1989, specsurveybartocci}.

Signal Temporal Logic (STL)~\cite{maler_monitoring_2004} provides a formal framework for specifying both timing constraints and state-dependent properties, making it well suited for verifying dynamical behaviors. For systems with modeling and disturbance uncertainties, stochastic methods are commonly used, extending single-trajectory temporal logic verification to a probabilistic setting as in~\cite{SankarMonteCarlo}. However, these methods may miss rare or worst-case behaviors, thus offering limited safety guarantees. To overcome this limitation, we adopt reachability analysis, which systematically over-approximates all possible system trajectories under bounded disturbances~\cite{alexandre_dit_sandretto_validated_2016, Althoff2015ARCH, capdarticle}. This approach ensures that no behavior is overlooked, enabling sound verification of STL specifications across both temporal and state domains.

In reachability analysis frameworks such as~\cite{alexandre_dit_sandretto_validated_2016, le_coent_improved_2018}, adaptive step size selection is employed to satisfy precision requirements. Although smaller step sizes may increase computational cost and introduce conservatism, they are sometimes essential to reduce the uncertainty in STL satisfaction. Nonetheless, given that the objective of this work is to verify temporal properties through STL specifications, any uncertainty in the satisfaction must stem from the intrinsic system dynamics, rather than being an artifact of the numerical step size selection.


\textit{Related Work.} Various approaches have been proposed for verifying STL formulas against reachable sets \cite{Lercher2024CAV, Roehm2016ATVA, Tillet2025ActaCybernetica, Ishii2016IEICE, Yu2024Automatica} or interval of signals~\cite{Baird2023IEEE}.  
Some proposed a new formalism known as Reachset Temporal Logic (RTL) providing a sound and complete transformation from an STL formula to an RTL formula \cite{Roehm2016ATVA} and an automatic refinement strategy in~\cite{kochdumper2024fully}. However, this approach constrains the time step and time range of the formula.
In Ishii et al. \cite{Ishii2016IEICE}, the concept of consistent time intervals is introduced, where the reachable set allows for a definitive determination of whether an STL formula is satisfied. These approaches fall under offline monitoring \cite{maler_monitoring_2004,Roehm2016ATVA, Ishii2016IEICE}, as it requires the prior computation of the system evolution before verifying the formula. Conversely, Lercher et al. \cite{Lercher2024CAV} propose an online monitoring approach, which incrementally computes only the necessary reachable sets for verification, thereby improving efficiency in long horizon system.
%
To address uncertainties in satisfaction, three-valued semantics for temporal logic were introduced in~\cite{Tillet2025ActaCybernetica, Ishii2016IEICE, Wright2020RV, monitorltlabate, Bauer2011TOSEM}, while four-valued satisfaction was proposed in~\cite{Lercher2024CAV} to differentiate between incomplete observations and intrinsic system ambiguity.
To mitigate the uncertainty caused by large integration step sizes, the approach in~\cite{Lercher2024CAV} employs smaller step sizes, potentially requiring multiple reachable set computations before evaluating formula satisfaction.

In this paper, we build upon these concept by introducing a method based on Boolean interval arithmetic~\cite{kearfott_logical_2019, Tillet2025ActaCybernetica} for evaluating STL formulas over reachable sets. Our approach supports the verification of nested STL formulas by using satisfaction signal, not always supported in prior works~\cite{Tillet2025ActaCybernetica, Yu2024Automatica}.
We decompose the satisfaction signal into certain and uncertain unitary Boolean signals, enabling uncertainty tracking within the satisfaction tree. Our approach identifies whether an uncertainty arises from specific reachable sets or from incomplete simulation, and selectively contracts or refines only the relevant sets. This method is orthogonal to approaches where verification is static on the generated sequence of reachable sets \cite{Wright2020RV, Ishii2016IEICE, Roehm2016ATVA, Tillet2025ActaCybernetica}. We assume access to a model of the system dynamics, allowing us to compute and contract reachable sets over specified time intervals.
Our automatic refinement approach also addresses the over-approximation of reachable sets, similar to~\cite{kochdumper2024fully}, but remains fully within the STL formalism and extends to nonlinear systems. In contrast to~\cite{kochdumper2024fully}, our method is designed for online monitoring under incomplete time horizons. By refining reachable sets through localized backtracking and parsimonious forward computation, it enables efficient verification, in line with the incremental strategy of~\cite{Lercher2024CAV}.
%

Our main contribution is a dynamic STL verification framework that reduces uncertainty by selectively refining only the reachable sets responsible for indeterminate satisfaction. The approach combines offline and online monitoring to improve precision while avoiding unnecessary computations.

Initially, Section~\ref{sec:Preliminaries} introduces the key concepts used throughout the paper. In Section~\ref{sec:stlsatisfset}, we extend STL satisfaction on reachable sets using Boolean interval arithmetic and present a decomposition of the satisfaction signal for uncertainty tracking. Section~\ref{sec:uncertaintytracking} details the tracking of uncertainty to the predicate level within the satisfaction tree, enabling identification of its origin from specific reachable sets.
In Section~\ref{sec:adaptativesampling}, we redefine time intervals and contract the relevant reachable sets. We finally present the proposed algorithm and an experiment illustrating the efficiency of our approach.

\section{Preliminaries}
\label{sec:Preliminaries}
\subsection{Reachability analysis}
\label{sec:Reachability Analysis}
Let us consider a continuous dynamical system modeled by a differential equation of the form:
\begin{equation}
\dot{y}(t) = f(y(t), w(t)), \quad y(t) \in \mathbb{R}^n,\ w(t) \in \mathcal{W},
\end{equation}
where $y(t)$ denotes the system state, $w(t)$ is a bounded external input, $\mathcal{W} \subseteq \mathbb{R}^p$ is a compact set. A function $f$ is a nonlinear function, such that for any initial state $y_0 \in \mathbb{R}$ and any
measurable input signal $w : \mathbb{R}^+ \to W$, the system admits a unique trajectory in $\mathbb{R}^n$ denoted by $\xi(\cdot, y_0, w)$.

\begin{definition}
The \textit{reachable set} at time $t \in \mathbb{R}^+$ from a set of initial states $\mathcal{Y}_0 \subseteq \mathbb{R}^n$ is defined as~\cite{alexandre_dit_sandretto_validated_2016, Althoff2015ARCH}:
\begin{equation}
\small
\text{Reach}_t(\mathcal{Y}_0, \mathcal{W}) = \left\{ \xi(t, y_0, w) \mid y_0 \in \mathcal{Y}_0,\ w(s) \in \mathcal{W},\ \forall s \in [0, t] \right\}
\end{equation}
\end{definition}
\begin{definition} The \textit{reachable tube} over a time interval $[0, T] \subseteq \mathbb{R}^+$ is given by~\cite{alexandre_dit_sandretto_validated_2016, Althoff2015ARCH}:
\begin{equation}
\text{Reach}_{[0,T]}(\mathcal{Y}_0, \mathcal{W}) = \bigcup_{s \in [0,T]} \text{Reach}_s(\mathcal{Y}_0 , \mathcal{W}).
\end{equation}
\end{definition}

In the process of solving an Initial Value Problem for a differential equation using validated numerical integration methods~\cite{alexandre_dit_sandretto_validated_2016}, an enclosure \([\tilde{y}_j]\) of the solution is computed for each time interval \([t_j, t_{j+1}]\), where \(t_{j+1} = t_j + h_j\) and \(h_j\) is the integration step size. This enclosure satisfies:
\begin{equation}
\text{Reach}_{[t_j,t_{j+1}]} \subseteq [\tilde{y}_j],\text{ with }
\bigcup_{j=0}^{N-1} [t_j, t_{j+1}] = [t_0, T],
\end{equation}
with $t_0$ the start time of the simulation and $T$ the end time. 

The set of trajectories over the interval $[t_0, T]$ captures all possible executions of the system. This continuous-time representation, referred to as a \textit{tube} and denoted $[\tilde{y}](t)$ for $t \in [t_0, T]$, is essential to preserve system behavior that may be lost through sparse time sampling.
 This sequence of reachable sets is the object of our analysis for STL formula satisfaction.
\subsection{Signal temporal logic}
\label{sec:STL}
STL is an extension of Linear Temporal Logic designed to specify properties of continuous-time systems~\cite{maler_monitoring_2004,specsurveybartocci}, for robotics and control applications.
STL formulas enable the expression of temporal properties in terms of time bounds, combining logical operators with bounded temporal operators. The syntax of STL is defined recursively as follows:
\begin{equation}
\phi := \mu \mid T \mid \neg \phi \mid \phi_1 \vee \phi_2 \mid \phi_1 \, U_{[a,b]} \, \phi_2,
\end{equation}
where $\phi$ is an STL formula, $T$ denotes the tautology, $\neg$ the negation, and $\mu$ is an atomic predicate specifying signal constraints (e.g., $x > 3$)~\cite{maler_monitoring_2004}.

The semantics is the following:
\begin{equation}
\begin{aligned}
(s, t) \models \mu &\iff \pi_\mu(s)[t] = T \\
(s, t) \models \neg \phi &\iff (s, t) \not\models \phi \\
(s, t) \models \phi_1 \lor \phi_2 &\iff (s, t) \models \phi_1 \; \text{or} \; (s, t) \models \phi_2 \\
(s, t) \models \phi_1 U_{[a,b]} \phi_2 &\iff \exists t' \in [t + a, t + b], \; (s, t') \models \phi_2 \\
& \quad \quad \quad \text{and} \; \forall t'' \in [t, t'], \; (s, t'') \models \phi_1 \\
F_{[a,b]} \phi = T U_{[a,b]}& \phi, \quad G_{[a,b]} \phi = \neg(F_{[a,b]} \neg\phi),
\end{aligned}
\label{semantiquestl}
\end{equation}
where $(s,t)$ represents the trace of a signal $s$ at time $t$, and $\pi_\mu(s)$ is a function that evaluates a predicate on the signal. 
$U_{[a,b]}$ is the ``until'' operator bounded by the time interval $[a, b]$. 
The semantic is interpreted as: "if a formula $\phi_2$ is satisfied at a time $t'\in [t+a,t+b]$, then another formula $\phi_1$ must be true from the evaluating time $t$ to the time $t'$~".
The Finally operator $F_{[a,b]}$ expresses the existence of a satisfaction within a time interval. The Globally operator $G_{[a,b]}$ specifies that the subformula must hold during the entire time interval. 
Other logical operators such as $\land$ and $\implies$ are defined using standard logical equivalences:
\begin{equation}
\phi_1 \land \phi_2 \equiv \neg (\neg \phi_1 \lor \neg \phi_2), \quad 
\phi_1 \implies \phi_2 \equiv \neg \phi_1 \lor \phi_2.
\end{equation}
\subsection{Satisfaction signals}
\label{sec:satisfactionsignals}
%
To unify predicate evaluation and temporal reasoning, the concept of satisfaction signal $\mathcal{S}_{\phi}(t)$ is defined in Eq.~\eqref{eq:truth_signal builder} from~\cite{maler_monitoring_2004,Lercher2024CAV}. This representation tracks the time-dependent satisfaction of each nested subformula, allowing Boolean operations such as $\vee$, $\land$ and $\neg$ to be applied directly to these signals.
\begin{definition}[satisfaction signal~\cite{maler_monitoring_2004,Lercher2024CAV}]
\begin{equation}
\begin{aligned}
    \mathcal{S}_{\phi}: \mathbb{R}^+ \to \mathbb{B},~
    t &\mapsto \begin{cases}
    1 & \quad \text{if } (s, t) \models \phi_i\\
    0 & \quad \text{if } (s, t) \not \models \phi_i\\
    \end{cases}
\end{aligned}.
\label{eq:truth_signal builder}
\end{equation}
\end{definition}

As defined in~\cite{maler_monitoring_2004}, the satisfaction signal of the Until operator is constructed using \textit{unitary} Boolean signals. A satisfaction signal $\mathcal{S}^*_{\phi}$ is said to be \textbf{unitary} if it is true only over a single, continuous and positive time interval $I_\phi$, denoted unitary time interval in this paper. This is a key concept as it allows for the verification of any STL formula with temporal operator.

\begin{figure}[htbp]
       \begin{center}
        \includegraphics[width=0.8 \linewidth]{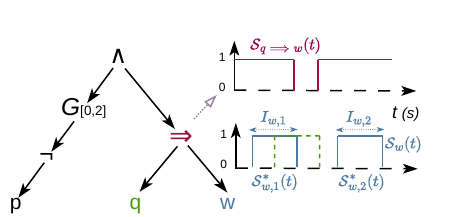}
      \caption{Satisfaction tree~\cite{maler_monitoring_2004} and associated satisfaction signals.}
      \label{sattree}
      \end{center}
   \end{figure}

The satisfaction signal of any formula can be given by a disjunction of $N$ unitary signals~\cite{maler_monitoring_2004}. The decomposition in unitary signals is illustrated in Fig.~\ref{sattree}. $\mathcal{S}_{w}(\cdot)$ is the satisfaction signal of the predicate $w$. It is decomposed in two unitary Boolean signals: $\mathcal{S}_{w,1}^*(\cdot)$ and $\mathcal{S}_{w,2}^*(\cdot)$
\begin{definition}[unitary decomposition~\cite{maler_monitoring_2004}]
\begin{equation}
    \mathcal{S}_{\phi}(t) := \bigvee_{j=0}^{N} \mathcal{S}^*_{\phi,j}(t),~t\in \mathbb{R^+}.
    \label{unitarybool}
\end{equation}
\end{definition}

For simplicity in notation, $\mathcal{S}_{\phi}(t), \forall t \in I$ corresponds to $\mathcal{S}_{\phi}(I)$. 
\begin{definition}
Following \cite{maler_monitoring_2004}, let $I = [m, n)$ and $[a, b]$ be intervals in $\mathbb{T} = \mathbb{R}^+$. The $[a, b]$-back shifting of $I$ is defined as $I \ominus [a, b] = [m - b, n - a) \cap \mathbb{T}$. 
\end{definition}

Consider two non-unitary satisfaction signals, $\mathcal{S}_{\phi_1}(t) = \bigvee_{i=1}^{n} \mathcal{S}_{\phi_1, i}^{*}(t)$ and $\mathcal{S}_{\phi_2}(t) = \bigvee_{j=1}^{m} \mathcal{S}_{\phi_2, j}^{*}(t)$, each composed of $n$, $m$ unitary signals with corresponding unitary intervals $I_{\phi_1,i}$ and $I_{\phi_2,j}$,
%
The satisfaction signal of $\psi = \phi_1 \mathcal{U}_{[a,b]} \phi_2$ is defined on multiple unitary time interval $I_{\psi, i,j}$ as follows~\cite{maler_monitoring_2004}.
%
%
%
\begin{lemma}[unitary decomposition of $U_{[a,b]}$~\cite{maler_monitoring_2004}]
\begin{equation}
    \begin{aligned}
        & \mathcal{S}_{\psi}(t) := \bigvee_{i=1}^{m} \bigvee_{j=1}^{n} \mathcal{S}_{\psi, i, j}^{*} (t), ~t\in \mathbb{T} \text{ with,} \\
      & I_{\psi, i,j} = \left( (I_{\phi_1,i} \cap I_{\phi_2,j}) \ominus [a,b] \right) \cap I_{\phi_1,i}, \text{ and }\\
      & \mathcal{S}_{\psi,i,j}^*(t) := \begin{cases}
          &  \mathcal{S}_{\psi,i,j}^{*} (I_{\psi,i,j})=1 \\
          &  \mathcal{S}_{\psi,i,j}^{*}(\mathbb{T} \setminus I_{\psi,i,j})=0
      \end{cases}.
    \end{aligned}
      \label{untilsatisfsignal}
\end{equation}
\end{lemma}

\vspace{4px}
In the context of offline monitoring, we look for the satisfaction of an STL formula. To achieve this, we recursively construct satisfaction signals in a bottom-up manner, starting from a satisfaction signal on predicates and proceeding until the ``root'' formula. This approach enables tracking the temporal properties of the signal.
Here is an example of an STL formula: $\phi = G_{[0,2]}(\neg p) \land (q \implies w)$, where $p$, $q$ and $w$ are predicates. Figure~\ref{sattree} represents the satisfaction tree. 
\section{STL satisfaction on reachable sets}
\label{sec:stlsatisfset}
The reachable tube, denoted $([\tilde{y}], t)$, encloses all system behaviors. The STL formalism, originally defined over a single trace, is extended to evaluate satisfaction over this set of traces.
\subsection{Boolean intervals}
\label{sec:booleaninterval}
While considering sets, constraints may be satisfied for some executions and violated for others, so the result cannot be simply true or false. To handle this, Boolean intervals are used, which extend traditional Boolean logic to manage uncertainties in set-based contexts~\cite{kearfott_logical_2019}. A Boolean interval is a subset of $\mathbb{B} = \{ 1, 0 \}$, belonging to the interval Boolean set $\mathbb{IB} = \{\emptyset, [0,0], [1,1], [0,1]\}$.
Here, the values $\emptyset$ and $[0,1]$ represent, respectively, impossible and undetermined states. In this paper, $[1,1]$ will be denoted as $1$ (true value), and $[0,0]$ as $0$ (false). The logical operations on Boolean values are extended to Boolean intervals $[a]$ and $[b]$ as follows.
\begin{definition}[Boolean interval arithmetic~\cite{kearfott_logical_2019, jaulin_applied_2001}]
\begin{equation}
  \begin{aligned}
[a] \wedge [b] &= \{a \wedge b \mid a \in [a], b \in [b]\}\\
[a] \vee [b] &= \{a \vee b \mid a \in [a], b \in [b]\}\\
\neg[a] &= \{\neg a \mid a \in [a]\}.
\end{aligned}
\label{booleanitvarithmetic}
\end{equation}
\end{definition}

For example, the behavior of an interval $[0,1]$ is:
\begin{align*}
0 \wedge [0,1] &= 0,\quad 0 \vee [0,1] = [0,1],\\
1 \wedge [0,1] &= [0,1],\quad 1 \vee [0,1] = 1.
\end{align*}

\subsection{Satisfaction on set predicates}
\label{sec:satisfactionsetp}
To extend predicate evaluation to a set-based context, two set predicates are defined from~\cite{Tillet2025ActaCybernetica, Lercher2024CAV}, the inclusion predicate $\mu_i$, and the exclusion predicate $\mu_d$, corresponding to, $y(t) \in \mathcal{X}^\mu$, and $y(t) \not\in \mathcal{X}^\mu$, respectively.

Here, $\mathcal{X}^\mu \in \mathbb{IR}^n$ is a set on an $n$-dimensional space, and $y(t) \in \mathbb{R}^n$ is the state of the system at time $t$. Abstracting sets by using level sets, this new syntax is not less general than previously.
For example, a predicate \(\mu = x>0\) is represented by an inclusion in a set \(\mathcal{X}^\mu=\{x\in\mathbb{R} \mid x>0\}\). Moreover, these predicates can be time-varying, provided that an exact model of their dynamics is available.
In practice, the introduced predicates are redundant but offer flexibility while defining constraints on the system.

\begin{definition}[set-based extension of predicate~\cite{Tillet2025ActaCybernetica, Lercher2024CAV}]
For the inclusion predicate:
\begin{equation}
  ([{{\tilde{y}}}], t) \vDash \mu _i :=
  \begin{cases}
    1,     & \text{if } [\tilde{{y}}](t) \subset \mathcal{X}^\mu, t \in [t_0, T],\\
    0,     & \text{if } [\tilde{{y}}](t) \cap \mathcal{X}^\mu = \emptyset, t \in [t_0, T],\\
    [0,1], & \text{otherwise.}
  \end{cases}
  \label{eq:set_predicate}
\end{equation}
\end{definition}
At the formula level, the following equivalence is established:
\begin{equation}
({{y}},t) \vDash \neg \mu_{i} \equiv ({{y}},t) \vDash \mu_{d},
\end{equation}
with $(y,t)$ being an execution of our system.

If the predicate cannot be conclusively evaluated over the reachable set $[\tilde{y}_j]$ on $[t_j, t_{j+1}]$, the satisfaction is marked as undetermined and the value $[0,1]$ is assigned.
Lastly, the set-membership STL syntax is written as:
\begin{equation}
  \phi := \mu_i \mid T \mid \lnot \phi \mid \phi_1 \lor \phi_2 \mid \phi_1\ \mathcal{U}_{[a,b]} \phi_2.
\end{equation}
\subsection{Unitary decomposition in uncertain context}
\label{sec:unitarydecomp}
As stated in Section~\ref{sec:satisfactionsignals}, handling temporal operators requires decomposing the satisfaction signal into unitary signals. Following the formalism introduced in Eq.~(\ref{unitarybool}), we define two unitary decompositions that are essential for evaluating the $U_{[a,b]}$ operator in uncertain context:
\begin{definition}[uncertain and certain unitary signals]
\begin{equation}
   \; \overline{ \mathcal{S}_{\phi}^{*}}:\mathbb{R}^+ \to \mathbb{IB},~t\mapsto
\begin{cases}
    [0,1] & \text{if } t \in \overline{I_\phi} \\
    0 & \text{if } t \in \mathbb{T} \setminus \overline{I_\phi}
\end{cases} 
\label{unitaryitvboolup}
\end{equation}
\begin{equation}
  \quad \quad \; \underline{ \mathcal{S}_{\phi}^{*}}:\mathbb{R}^+ \to \mathbb{IB},~t\mapsto
\begin{cases}
    1 & \text{if } t \in \underline{I_\phi} = 1 \qquad \quad \\
    0 & \text{if } t \in \mathbb{T} \setminus \underline{I_\phi},
\end{cases} 
\label{unitaryitvbooldown}
\end{equation}
where $\overline{I_\phi}$ denotes unitary time intervals with only uncertain satisfaction $[0,1]$, and $\underline{I_\phi}$ those with guaranteed satisfaction~$1$.
\end{definition}

%
Finally, we define the satisfaction signal at $t$ by:
\begin{definition}[uncertain unitary decomposition]
\begin{equation}
    [\mathcal{S}_{\phi}]:\mathbb{R}^+ \to \mathbb{IB},~t\mapsto \bigvee_{j=0}^{N} \overline{ \mathcal{S}_{\phi, j}^{*}}(t) \vee \bigvee_{i=0}^{K} \underline{ \mathcal{S}_{\phi, i}^{*}}(t).
    \label{unitaryitvbool}
\end{equation}
\end{definition}
An illustrative example is provided in Fig.~\ref{tracking logic}.
This decomposition captures how satisfaction is composed. The satisfied portions of the signal are computed as $\bigvee_{i=0}^{K} \underline{ \mathcal{S}_{\phi, i}^{*}}(\cdot)$, consistent with the construction in~\cite{maler_monitoring_2004}, while $\bigvee_{j=0}^{N} \overline{ \mathcal{S}_{\phi, j}^{*}}(\cdot)$ captures intervals where satisfaction remains uncertain. Lastly, portions of $[\mathcal{S}_{\phi}](\cdot)$ with a satisfaction value of $0$ correspond to guaranteed violations.

If at time $t$, $\underline{ \mathcal{S}_{\phi, i}^{*}}(t) = 0$ and $\overline{ \mathcal{S}_{\phi, i}^{*}}(t) = [0,1]$, then the Boolean interval arithmetic, defined in Eq.~\eqref{booleanitvarithmetic}, gives $0 \lor [0,1] = [0,1]$, indicating that the signal may still be satisfied. Conversely, if $\underline{ \mathcal{S}_{\phi, i}^{*}}(t) = 1$ and $\overline{ \mathcal{S}_{\phi, i}^{*}}(t) = [0,1]$, then $1 \lor [0,1] = 1$, meaning that the satisfaction is definitively true, and the uncertainty has no further impact—thus maintaining consistency with the semantics from~\cite{maler_monitoring_2004} in Eq.~\eqref{unitarybool}.

Finally, Boolean operations on satisfaction signals are extended following the arithmetic introduced in~Eq. \eqref{booleanitvarithmetic}:
    \begin{definition}[$\lor$ operator on interval satisfaction signal]
\begin{equation}
     \quad \quad \; \quad [\mathcal{S}_{\phi_1 \vee \phi_2}](t) := \begin{cases}
         & 1 ,\text{ if } [\mathcal{S}_{\phi_1}](t) \vee [\mathcal{S}_{\phi_2}](t) = 1 \\
         & 0 ,\text{ if } [\mathcal{S}_{\phi_1}](t) \vee [\mathcal{S}_{\phi_2}](t) = 0\\
         & [0,1] ,\text{ otherwise.}
     \end{cases}
\end{equation}
\end{definition}
\begin{definition}[$\neg$ operator on interval satisfaction signal]
\begin{equation}
     [\mathcal{S}_{\neg \phi_1}](t) := \begin{cases}
         & 1 ,\text{ if } [\mathcal{S}_{\phi_1}](t) = 0 \\
         & 0 ,\text{ if } [\mathcal{S}_{\phi_1}](t) = 1\\
         & [0,1] , \text{ otherwise.}
     \end{cases}
\end{equation}
\end{definition}
%
%
%
\subsection{Computing unitary signals from predicates}
\label{sec:boolpredicate}
To compute the overall formula satisfaction, it starts at the predicate level by identifying two unitary signals. The unitary time interval $\underline{I_\mu}$ and $\overline{I_\mu}$ are constructed by merging consecutive time intervals $[t_j, t_{j+1}]$ where the predicate is consistently satisfied, i.e., $\left( \bigcup_{j_0}^{j_f} [\tilde{y}_j] \right) \subset \mathcal{X}^\mu$, or consistently uncertain, i.e., $\left( \bigcup_{j_0}^{j_f} [\tilde{y}_j] \cap \mathcal{X}^\mu \right) \ne \emptyset$ and $\left( \bigcup_{j_0}^{j_f} [\tilde{y}_j] \right) \not \subset \mathcal{X}^\mu$. This forms a merged unitary time interval $I_\mu = [t_{j_0}, t_{j_f}]$.

Finally, we compute the satisfaction signal of a predicate:
\begin{equation}
\begin{aligned}
    &[\mathcal{S}_{\mu}](t) := \bigvee_{j=0}^{N} \overline{ \mathcal{S}_{\mu, j}^{*}}(t) \vee \bigvee_{i=0}^{K} \underline{ \mathcal{S}_{\mu, i}^{*}}(t) \text{, with} \\
    & \overline{ \mathcal{S}_{\mu}^{*}}(t) :=
   \begin{cases}
      [0,1], & \text{if }  t \in \overline{I_\mu} \\
      0, & \text{if } t \in \mathbb{T} \setminus \overline{I_\mu}
   \end{cases} \\
   & \underline{ \mathcal{S}_{\mu}^{*}}(t) := 
   \begin{cases}
      1, & \quad \; \; \text{if }  t \in \underline{I_\mu} \\
      0, &  \quad \; \; \text{if } t \in \mathbb{T} \setminus \underline{I_\mu}.
   \end{cases}
\end{aligned}
\label{unitaryitvboolmu}
\end{equation}
\subsection{Until operator under uncertainty}
\label{sec:Untilunitarybool}
The definition in Eq.~\eqref{untilsatisfsignal} provides a unitary Boolean decomposition of the Until operator, as proposed by Maler. While this decomposition allows us to trace uncertainties within the satisfaction tree, it was originally formulated for a deterministic setting and thus requires adaptation to account for uncertainty.

%
Building on the satisfaction semantics of Until defined by~\cite{Lercher2024CAV}, the Until satisfaction is as follows:
\begin{definition}[$U_{[a,b]}$ operator in uncertain context~\cite{Lercher2024CAV}]
\begin{equation}
\begin{aligned}
  &([{{\tilde{y}}}], t) \vDash \phi_1 U_{[a,b]} \phi_2 := \\ 
  &\begin{cases}
    1,  & \text{if } \exists t' \in [t + a, t + b],~ [\mathcal{S}_{\phi_2}](t')=1 \\ 
        & \quad \text{and } \forall t'' \in [t, t'], \; [\mathcal{S}_{\phi_1}](t'')=1, \\
    0,  & \text{if } \forall t' \in [t+a, t+b], ~ [\mathcal{S}_{\phi_2}](t')=0 \\ 
        & \quad \text{or } \exists t'' \in [t, t'],~ [\mathcal{S}_{\phi_1}](t'')=0, \\
    [0,1], & \text{otherwise}.
  \end{cases}
\end{aligned}
\label{eq:guaranteeduntil}
\end{equation}
\end{definition}

A unitary decomposition for the Until operator, consistent with the structure defined in Eq.~\eqref{unitaryitvbool}, is proposed. 
We define four unitary signals with corresponding unitary time intervals $\overline{I'_{\phi_1}}$, $\underline{I_{\phi_1}}$, $\overline{I_{\phi_2}}$ and $\underline{I_{\phi_2}}$ defined on positive times with $\overline{ \mathcal{S}_{\phi_1}^{*}}(\cdot)$, $\underline{ \mathcal{S}_{\phi_1}^{*}}(\cdot)$, $\overline{ \mathcal{S}_{\phi_2}^{*}}(\cdot)$ and $\underline{ \mathcal{S}_{\phi_2}^{*}}(\cdot)$. 
%
%
As stated in the definition of $U_{[a,b]}$, the subformula $\phi_1$ must hold over the entire interval $[t, t']$. If $\phi_1$ holds everywhere except at some $t'' \in [t, t']$ where $\mathcal{S}_{\phi_1}(t'') = [0,1]$, then the overall satisfaction of $\phi_1$ is conservatively set to $[0,1]$. However, this uncertainty may be resolved to $1$ or $0$ through further refinement or additional computations.

\begin{figure}[htbp]
    \begin{center}
        \includegraphics[width=\linewidth, trim=10 3 75 0, clip]{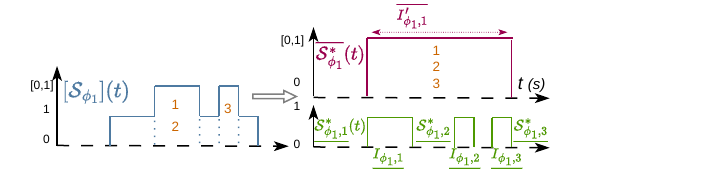}
        \caption{Unitary signal decomposition of $[\mathcal{S}_{\phi_1}](t)$ in $\phi_1 U_{[a,b]} \phi_2$; uncertainty indices in orange.}
        \label{unitary_decomposition_logic}
    \end{center}
\end{figure}

Then the definition of the unitary time interval $\overline{I_{\phi_1}}$ is different from Eq.~\eqref{unitaryitvboolup} in this approach. Considering a new unitary time interval $\overline{I'_{\phi_1}}$, it is defined as: $\forall t \in \overline{I'_{\phi_1}}, [\mathcal{S}_{\phi_1}](t) = 1 \text{ or } [\mathcal{S}_{\phi_1}](t) = [0,1]$.
An illustration is proposed in Fig.~\ref{unitary_decomposition_logic}.
\begin{equation}
   \overline{ \mathcal{S}_{\phi_1}^{*}}(t) := 
\begin{cases}
    [0,1] & \text{if } t \in \overline{I'_{\phi_1}},\\
    0, & \text{if } t \in \mathbb{T} \setminus \overline{I'_{\phi_1}}.
\end{cases} 
\label{unitaryitvboolphi1}
\end{equation}

Finally, we propose the interval satisfaction signal of Until as follows.
\begin{proposition}[satisfaction signal of $\psi = \phi_1 \mathcal{U}_{[a,b]} \phi_2$]
\noindent
\begin{equation}
    [\mathcal{S}_{\psi}](t) := \underline{\mathcal{S}_{\psi}}(t) \vee \overline{\mathcal{S}_{\psi}}(t),
\end{equation}
\noindent
with the signal $\overline{ \mathcal{S}_{\psi}}(\cdot)$:
\begin{equation}
\begin{aligned}
    & \overline{\mathcal{S}_{\psi}}(t) = \overline{\mathcal{S}_{\psi}}_{\underline{\phi_2}}(t) \vee \overline{\mathcal{S}_{\psi}}_{\overline{\phi_2}}(t), \text{ with, }\\
   &\begin{cases}
           & \overline{\mathcal{S}_{\psi}}_{\underline{\phi_2}}(t) = \bigvee_{i=1}^{m} \bigvee_{j=1}^{n}\overline{\mathcal{S}_{\psi}^{*}}_{\underline{\phi_2},i,j}(t),\\
           & \text{with, } \overline{I_{\psi}}_{\underline{\phi_2},i,j} = \left( \left( \overline{I'_{\phi_1,i}} \cap \underline{I_{\phi_2,j}} \right) \ominus [a,b] \right) \cap \overline{I'_{\phi_1,i}},
   \end{cases}\\
   &\text{and, }\\
   &\begin{cases}
            & \overline{\mathcal{S}_{\psi}}_{\overline{\phi_2}}(t) = \bigvee_{i=1}^{m} \bigvee_{j=1}^{n} \overline{\mathcal{S}_{\psi}^{*}}_{\overline{\phi_2},i,j}(t),\\
           & \text{with, } \overline{I_{\psi}}_{\overline{\phi_2},i,j} = \left( \left( \overline{I'_{\phi_1,i}} \cap \overline{I_{\phi_2,j}} \right) \ominus [a,b] \right) \cap \overline{I'_{\phi_1,i}},
   \end{cases}
\end{aligned}
\label{unitaryUntilitv}
\end{equation}
with the signal $\underline{ \mathcal{S}_{\psi}}(\cdot)$:
\begin{equation}
\begin{aligned}
    & \underline{\mathcal{S}_{\psi}}(t) = \bigvee_{i=1}^{m} \bigvee_{j=1}^{n} \underline{\mathcal{S}_{\psi,i,j}^{*}} \; ,  \text{ with, }\\
    & \underline{I_{\psi,i,j}} = \left( \left( \underline{I_{\phi_1,i}} \cap \underline{I_{\phi_2,j}} \right) \ominus [a,b] \right) \cap \underline{I_{\phi_1,i}}. \\
\end{aligned}
\label{unitaryUntilitvlower}
\end{equation}
\end{proposition}
%
\paragraph*{Sketch of Proof}  
We need to prove the soundness of the decomposition $[\mathcal{S}_{\psi}](t) := \underline{\mathcal{S}_{\psi}}(t) \vee \overline{\mathcal{S}_{\psi}}(t)$.
Starting from the decomposition of unitary signals given in Eq.~\eqref{untilsatisfsignal}~\cite{maler_monitoring_2004}, our approach differs in the treatment of temporal quantifiers on $\phi_1$ and $\phi_2$. As previously defined, $\overline{I_{\phi_2}}$ or $\underline{I_{\phi_2}}$ indicates continuous satisfaction of $\phi_2$, but the existential quantifier $\exists t'$ removes dependency on successive unitary signals. Conversely, for $\phi_1$, the universal quantifier $\forall t \in [t, t']$ introduces dependencies between successive unitary signals, potentially transitioning as $1 \to [0,1] \to 1$.  

$\overline{\mathcal{S}_{\psi}}(\cdot)$ is composed by case distinction, aiming to capture time intervals with guaranteed $0$ satisfaction. Due to the universal quantifier on $\phi_1$, any combination of $1$ and $[0,1]$ over $\overline{I'_{\phi_1}}$ yields $[0,1]$, but never $0$, justifying the semantics of $\overline{\mathcal{S}_{\phi_1}^{*}}(\cdot)$. 
Specifically, $\overline{I_{\psi}}_{\underline{\phi_2}}$ defines a unitary time interval $\overline{I_\psi}$ of an uncertain Until unitary signal where $\phi_1$ is uncertain, but $\phi_2$ is certain. Similarly, $\overline{\mathcal{S}_{\psi}^{*}}_{\overline{\phi_2}}(\cdot)$ represents an uncertain Until unitary signal where both condition on $\phi_1$ and $\phi_2$ are uncertain. 
%

Additionally, in our semantics, the signal $\overline{ \mathcal{S}_{\phi_1}^{*}}(\cdot)$ already accounts for time intervals where the satisfaction is $1$. When intersecting the shorter unitary interval $\underline{I_{\phi_1}} \subset \overline{I'_{\phi_1}}$ with $\overline{I_{\phi_2}}$, the resulting interval $\overline{I_{\psi_{\underline{\phi_1},\overline{\phi_2}}}} = \left( \left( \underline{I_{\phi_1}} \cap \overline{I_{\phi_2}} \right) \ominus [a,b] \right) \cap \underline{I_{\phi_1}}$
is necessarily included in $\overline{I_{\psi}}_{\overline{\phi_2}}$, 
due to the semantics of back-shifting and intersection applied to these unitary intervals, moreover, the satisfaction of $\psi$ over $\overline{I_{\psi_{\underline{\phi_1},\overline{\phi_2}}}} \subset \overline{I_{\psi}}_{\overline{\phi_2}}$ remains $[0,1]$.

Finally, $\underline{\mathcal{S}_{\psi}}(\cdot)$ corresponds to a unitary signal with guaranteed $1$ satisfaction, consistent with the initial decomposition (Eq.~\eqref{untilsatisfsignal})~\cite{maler_monitoring_2004} and the semantics for Until in uncertain contexts (Eq.~\eqref{eq:guaranteeduntil})~\cite{Lercher2024CAV}.
Thus, $[\mathcal{S}_{\psi}](t) := \underline{\mathcal{S}_{\psi}}(t) \vee \overline{\mathcal{S}_{\psi}}(t)$ maintains the same logical composition as Eq.~\eqref{unitaryitvbool}, completing the proof.
%

%
\section{Uncertainty tracking in logical Operators}
\label{sec:uncertaintytracking}
Uncertainties in STL satisfaction may stem from incomplete simulation or over-approximated reachable sets. To distinguish these sources and enable targeted refinement, a method is introduced to track the origin of uncertainties through unitary signal decomposition.
Uncertain time intervals on predicates, represented by the corresponding reachable set indices, are memorized and propagated through the satisfaction tree (Section~\ref{sec:satisfactionsignals}) using simple rules as detailed in the following.
These memorized uncertain time intervals are called ``markers''. They point out the reachable sets that contribute to uncertainty in the final satisfaction.

\subsection{Tracking from predicate level to the root}
\label{sec:trackingpred}
Starting from Eq.~\eqref{unitaryitvbool} and Eq.~\eqref{eq:set_predicate}, if a reachable set $[\tilde{y}_j]$ defined on $[t_j, t_{j+1}]$ satisfies $[\tilde{y}_j] \cap \mathcal{X}^\mu \neq \emptyset$ and $[\tilde{y}_j] \not\subset \mathcal{X}^\mu$  as in Fig. \ref{contractedbox}, an uncertainty arises. As discussed in Section~\ref{sec:boolpredicate}, $\overline{I_\mu}$ is computed as the union of multiple time intervals $[t_j, t_{j+1}]$. If multiple reachable sets intersect the predicate, all corresponding indices $j$  are associated with this unitary signal $\overline{S_\mu^*}$ using a marker denoted $\mathcal{M}_{\overline{I_{\mu}}}$:
$$
\mathcal{M}_{\overline{I_{\mu}}} = \bigcup_{j_0}^{j_f} \{j\}, \quad \text{with } \overline{I_{\mu}} = [t_{j_0}, t_{j_f}].
$$

\subsection{Boolean operator}
\label{sec:trackingbool}
\begin{figure}[htbp]
       \begin{center}
        \includegraphics[width=\linewidth, trim=0 10 0 5, clip]{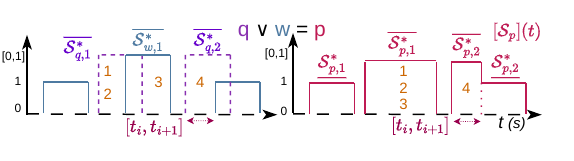}
      \caption{Uncertainty tracking in the $\vee$ operation between two satisfaction signal (in color) using unitary signals; uncertainty indices in orange.}
      \label{tracking logic}
      \end{center}
   \end{figure}

\paragraph{OR}
To track uncertainties, we retrieve the markers originating from the unitary signals of the subformulas. A logic is then established to extract and propagate these markers.
Practically, when computing new unitary signals (i.e. $\overline{S_{p,2}^*}$ in Fig.~\ref{tracking logic}), if a new unitary time interval $\overline{I} = [t_i, t_{i+1}]$ is associated with $[0,1]$, we assign a marker to this interval containing all uncertainty indices $j$ from reachable sets $[\tilde{y}_j]$ (i.e. coming from $\overline{S_{q,2}^*}$). Specifically, when computing $[0,1] \lor 1 = 1$, all markers from the $[0,1]$ satisfaction are discarded for the resulting time interval. Finally, identical satisfaction time intervals are merged to form unitary signals, incorporating all corresponding markers (i.e. $\overline{S_{p,1}^*}$ in Fig.~\ref{tracking logic}).

Let $\phi = \phi_1 \vee \phi_2$ be a formula. Given a newly computed unitary interval $\overline{I_{\phi}}$ and contributing unitary intervals $\overline{I_{\phi_1}}$ and $\overline{I_{\phi_2}}$, the marker is assigned as:
\[
\mathcal{M}_{\overline{I_{\phi}}} = 
\begin{cases}
\mathcal{M}_{\overline{I_{\phi_1}}}, & \text{if } [S_{\phi_2}](\overline{I_{\phi}}) = 0 \\
\mathcal{M}_{\overline{I_{\phi_2}}}, & \text{if } [S_{\phi_1}](\overline{I_{\phi}}) = 0 \\
\mathcal{M}_{\overline{I_{\phi_2}}} \cup \mathcal{M}_{\overline{I_{\phi_1}}}, & \text{if } \overline{I_{\phi_1}} \cap  \overline{I_{\phi_2}} \neq \emptyset.
\end{cases}
\]
   
\paragraph{NEG} Negation is straightforward: $(s, t) \models \neg \phi \iff (s, t) \not\models \phi$.  
Therefore, $\neg [\mathcal{S}_{\phi}](\cdot)$ corresponds to the negation of the satisfaction for each time interval in $[\mathcal{S}_{\phi}](\cdot)$. The markers are preserved, as the negation of an uncertain satisfaction remains uncertain.

\subsection{Temporal operators} 
\label{sec:trackingtemp}
\paragraph{Until}
Starting from Eq.~\eqref{unitaryitvboolphi1} with $\psi = \phi_1 \mathcal{U}_{[a,b]} \phi_2$, the computation of the new unitary signal $\overline{\mathcal{S}_{\phi_1,i}^{*}}(\cdot)$ is associated with the union of contributing markers, forming a new marker on $\overline{I'_{\phi_1,i}}$ (see Fig.~\ref{unitary_decomposition_logic}).

From Eq.~\eqref{unitaryUntilitv}, during the computation of the Until operator, successive back-shifting and intersections are applied over unitary time intervals $\overline{I'_{\phi_1,i}}$ and $\overline{I_{\phi_2,j}}$. Throughout this process, the corresponding markers $\mathcal{M}_{\overline{I'_{\phi_1,i}}}$ or $\mathcal{M}_{\overline{I_{\phi_2,j}}}$ are merged to form new markers $\mathcal{M}_{\psi_{\underline{\phi_2},i,j}}$ and $\mathcal{M}_{\psi_{\overline{\phi_2},i,j}}$ corresponding to the unitary signals $\overline{\mathcal{S}_{\psi_{\underline{\phi_2},i,j}}^*}$, $\overline{\mathcal{S}_{\psi_{\overline{\phi_2},i,j}}^*}$ respectively.

To compute $\overline{\mathcal{S}_{\psi}}(\cdot)$, the disjunction ($\vee$) of all resulting unitary signals $\overline{\mathcal{S}_{\psi}^{*}}(\cdot)$ is taken, with the uncertainty tracking mechanism from Section~\ref{sec:trackingbool} applied to mark each resulting unitary interval with its sources of uncertainty. This same tracking logic is used when combining with $\underline{\mathcal{S}_{\psi}}(\cdot)$ in $[\mathcal{S}_{\psi}](t) := \underline{\mathcal{S}_{\psi}}(t) \vee \overline{\mathcal{S}_{\psi}}(t),~ t \in \mathbb{T}$, ensuring consistent propagation of uncertainty throughout the evaluation.

\paragraph{Finally and Globally} 
Since the Until operator is handled, and the temporal operators $F_{[a,b]}$ and $G_{[a,b]}$ are derived from it as:
\[
F_{[a,b]} \phi = T\, U_{[a,b]} \phi, \quad G_{[a,b]} \phi = \neg(F_{[a,b]} \neg\phi),
\]
the tracking mechanism naturally extends to both operators.

%
\subsection{Handling incomplete simulation}
\label{sec:trackingincomplete}
Following~\cite{maler_monitoring_2004}, signal comparison relies on the minimal signal length defined by the formula semantics. To ensure conservativeness and enable further computation, shorter signals are extended with $[0,1]$ to the required length. This avoid truncation errors particularly in temporal operators like $U_{[a,b]}$, preserving information and preventing premature conclusions. 

This conservative logic maintains consistency, e.g., $[0,1] \land 1 = [0,1]$, $[0,1] \land 0 = 0$, $[0,1] \lor 1 = 1$.

To distinguish such completions, a specific marker with a common negative indices is assigned to $[0,1]$ values introduced by signal extension, separating them from uncertainty originating in reachable sets.
   \begin{figure}[bp]
       \begin{center}
        \includegraphics[width=0.6\linewidth, trim=0 0 65 0, clip]{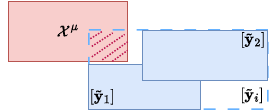}
      \caption{Time bisection and contraction, $\mathcal{X}_\mu$ a set predicate.}
      \label{contractedbox}
      \end{center}
   \end{figure}
\section{Adaptative sampling approach}
\label{sec:adaptativesampling}
\subsection{Recursive bisection using tracked error}
\label{sec:adaptativebis}
The step size in our approach is governed by a precision criterion. Specifically, validated numerical integration presented in Section~\ref{sec:Reachability Analysis}, dynamically adjusts the step size $h_j$ to satisfy a given precision constraint, which is closely related to the Local Truncation Error (LTE). This aligns with the approach in \cite{alexandre_dit_sandretto_validated_2016}, where LTE is computed using the order conditions of Runge-Kutta methods to ensure numerical stability and accuracy. 

However, excessively small step sizes not only increase computational cost but may also lead to overly pessimistic results, as the simulation becomes longer and the accumulated LTE may grow beyond what is necessary. To mitigate this, our approach balances step size selection to maintain accuracy while avoiding excessive computational overhead.

This causes two problems. First, some boxes may exhibit very large step times (in the context of dynamic equations), leading to uncertainty in transition times when strict timing conditions are considered. Second, these boxes may appear excessively large due to the need to cover the system state set over an extended time interval. This property introduces timing uncertainty and leads to overly conservative enclosures, potentially causing false crossing statements.

The overall refinement procedure is summarized in Algorithm~\ref{refinementalgo}. 
First, an initial tube is computed with \texttt{compute\_tube\_init()} (line~3). 
Then, an iterative loop progressively refines the tube until the STL satisfaction can be determined or the stopping criterion is met. 

At each iteration, the current tube is evaluated against the STL specification using 
\texttt{stl\_verifier()} (line~6). The resulting satisfaction signal, which may contain uncertain values, 
is analyzed by \texttt{uncertainty\_tracker()} (line~7), propagating uncertainty through the satisfaction tree and identifying the reachable sets that are responsible. 
If no uncertainty markers remain, the procedure terminates. Otherwise, the corresponding boxes are refined: 
their time intervals are bisected and the results are contracted using 
\texttt{bisect\_and\_contract()} (line~11). This refinement is performed by recomputing two new \emph{Picard boxes} with an additional integration scheme. 
The contraction step reduces as much as possible the diameter of the enclosures, 
thereby tightening the approximation of the trajectories~\cite{alexandre_dit_sandretto_validated_2016}. 
For example, a single box $[\tilde{\mathbf{y}}_i]$ is replaced with two smaller boxes $[\tilde{\mathbf{y}}_1]$ and $[\tilde{\mathbf{y}}_2]$, 
as illustrated in Fig.~\ref{contractedbox}. 

If the uncertainty originates from an incomplete simulation, detected via 
\texttt{exists\_negative\_index()} (line~12), the horizon is extended by $h$ and the missing portion of the tube is appended with 
\texttt{complete\_tube()} (line~14). 
This process repeats until all uncertainties are resolved or the stopping threshold is reached, 
and the algorithm finally returns the refined satisfaction.

\begin{algorithm}[hbp]
    \scriptsize
    \caption{Adaptive Refinement of a Tube under STL Satisfaction}
    \label{alg:adaptive_stl_verification}
    \begin{algorithmic}[1]
        \State \textbf{Input:} System parameters \texttt{params}, initial final time \texttt{final\_time}, STL \texttt{formula}, set of \texttt{predicates}, stopping threshold \texttt{stopping\_criterion}
        \State \textbf{Output:} Refined \texttt{satisfaction}

        \State \texttt{tube} $\gets$ \texttt{compute\_tube\_init}(\texttt{params}, \texttt{final\_time})
        \State \texttt{simulation\_complete} $\gets$ \texttt{False}

        \While{not \texttt{simulation\_complete}}
            \State \texttt{satisfaction} $\gets$ \texttt{stl\_verifier}(\texttt{predicates}, \texttt{formula}, \texttt{tube})
            \State \texttt{uncertainty\_marks} $\gets$ \texttt{uncertainty\_tracker}(\texttt{satisfaction}, \texttt{stopping\_criterion})

            \If{\texttt{uncertainty\_marks} is empty}
                \State \texttt{simulation\_complete} $\gets$ \texttt{True}
            \Else
                \State \texttt{tube} $\gets$ \texttt{bisect\_and\_contract}(\texttt{tube}, \texttt{uncertainty\_marks})
                
                \If{\texttt{exists\_negative\_index}(\texttt{uncertainty\_marks})}
                    \State \texttt{final\_time} $\gets$ \texttt{final\_time + h}
                    \State \texttt{tube} $\gets$ \texttt{complete\_tube}(\texttt{param}, \texttt{final\_time}, \texttt{tube})
                \EndIf
            \EndIf
        \EndWhile

        \State \Return \texttt{satisfaction}
    \end{algorithmic}
    \label{refinementalgo}
\end{algorithm}
%
%
%
\subsection{Case Study}
\label{sec:adaptativeCase}
\begin{figure}[ht]
    \includegraphics[width=\linewidth, trim=28 8 12 20, clip]{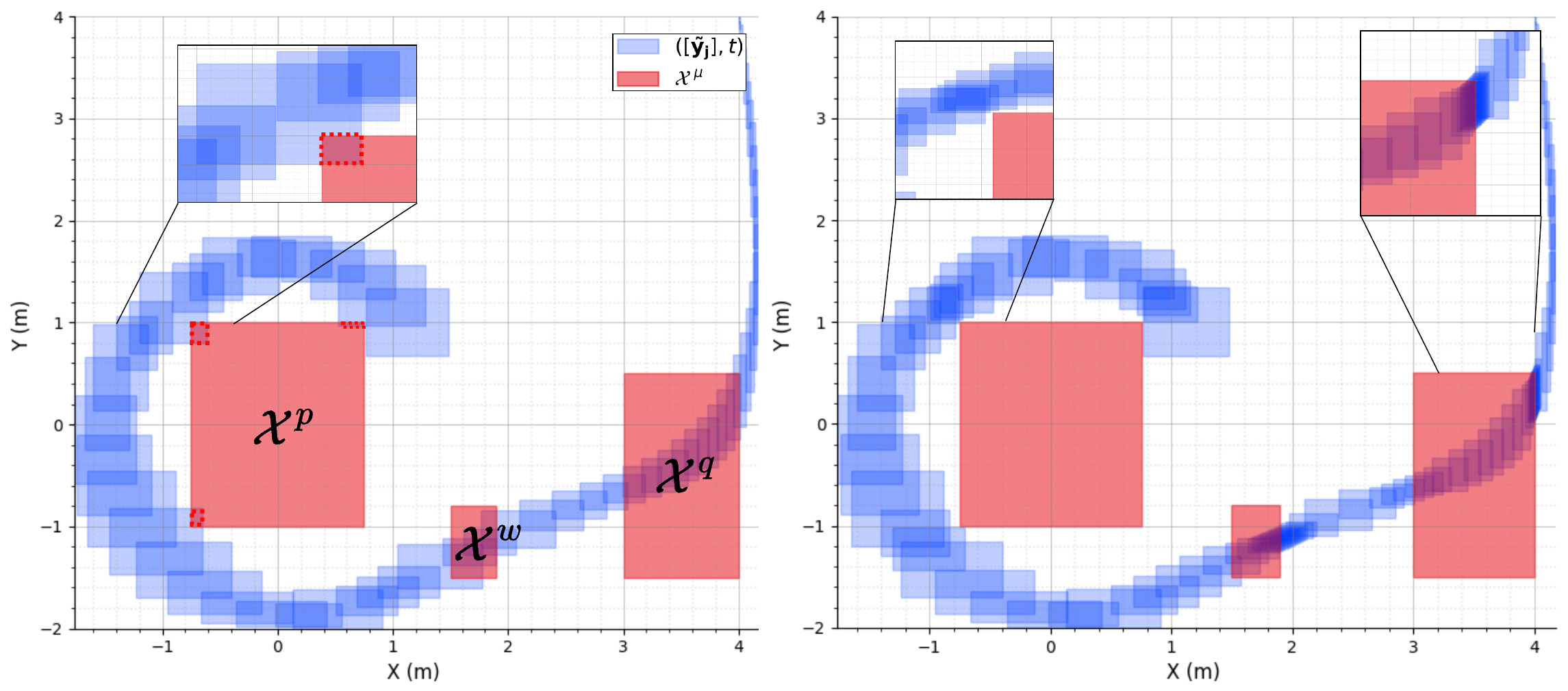}
    \caption{Initial tube on the left and refined tube on the right.}
    \label{beforeafter}
\end{figure}
In our approach, to perform reachability analysis, we are using a tool based on affine arithmetic and Runge-Kutta methods called DynIbex\footnote{https://perso.ensta-paris.fr/{{\raisebox{0.5ex}{\texttildelow}}}chapoutot/dynibex/}. Other tools, such as CORA\footnote{https://tumcps.github.io/CORA/}~\cite{Roehm2016ATVA, Lercher2024CAV} and CAPD\footnote{http://capd.ii.uj.edu.pl/}~\cite{capdarticle}, enable the reachability analysis of continuous dynamic systems and hybrid systems. The presented method generalizes to systems described by ODEs and any STL formula expressed with set predicates.

To demonstrate the efficiency of the proposed approach, an STL formula is verified on a tube, similar to the one studied in Section~\ref{sec:satisfactionsignals}: $\phi = G_{[0,1]}(G_{[0,2]}(\neg p) \land (q \implies F_{[1,2]} w))$. 
The tube dynamics are derived from a system governed by the Van der Pol oscillator and simulated over a 10 second horizon. Described as:
\begin{equation}
\begin{aligned}
\begin{cases}
    \dot{x} &= y, \\
    \dot{y} &= \mu (1 - x^2) y - x.
\end{cases}
\end{aligned}
\end{equation}
A comparison between the initial (left) and refined (right) tubes is shown in Fig.~\ref{beforeafter}.
The predicate $q$ is associated with the region on the right (red box), $\mathcal{X}^w$ with the central region, and $\mathcal{X}^p$ with the leftmost region. These correspond to spatial constraints in the STL specification. In natural language, this formula states that on a 1 second horizon, whenever the system encounters the $\mathcal{X}^q$ zone, it must reach the $\mathcal{X}^w$ zone within $[1,2]$ seconds. Additionally, within a $[0,2]$ time horizon, the system must never enter the $p$ box, which represents a forbidden zone.

As observed in the upper-left zoom, rediscretizing and contracting the boxes reduced pessimism and eliminated the uncertainty regarding the crossing of the forbidden zone. The upper-right zoom highlights a finer temporal resolution that mitigates the uncertainty on the moment of crossing $\mathcal{X}^q$, which enforces a temporal constraint in $\mathcal{X}^w$.

\begin{figure}[t!]
       \begin{center}
        \includegraphics[width=1\linewidth, trim=15 5 8 10, clip]{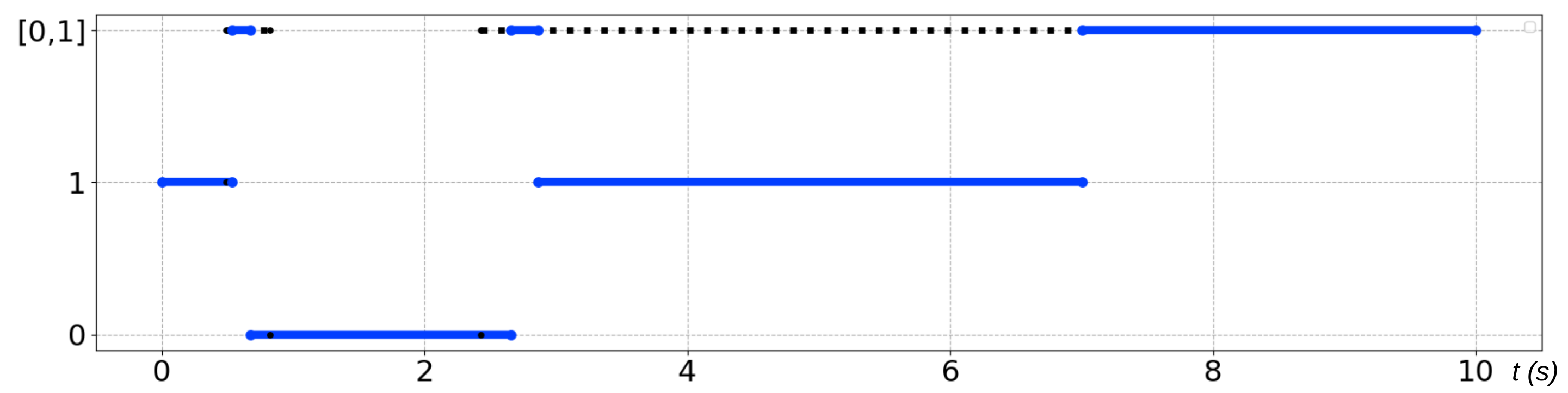}
      \caption{Original satisfaction signal $\mathcal{S}_\phi(t)$ in black dots vs. refined one in blue.}
      \label{satisfsignals}
      \end{center}
   \end{figure}

Finally, analyzing the satisfaction signals (Fig.~\ref{satisfsignals}) over a 10~s time horizon (simulation duration) demonstrates that, despite an initial 4.5~s uncertainty between 0~s and 7~s, our approach successfully mitigates uncertainty, reducing it to 0.5~s of uncertainty and 6.5~s of guaranteed satisfaction within this interval.
In this approach, the tube enables formula computation from 0~s to 7~s due to a 3~s shift (Fig.~\ref{satisfsignals}).
In our program, the remaining uncertainty $[0,1]$ from 7~s to 10~s is flagged as originating from an incomplete simulation.
\paragraph*{Computation Time}
On the same model and specification, we compared three approaches using the same hardware setup (an HP EliteBook with an Intel Core i7 processor): our uncertainty tracking mechanism implemented with DynIbex, an incremental approach inspired by~\cite{Lercher2024CAV} that we also implemented within DynIbex for a fair comparison, and the four-valued semantics approach using CORA~\cite{Lercher2024CAV}. The corresponding computation times are reported in Table~\ref{tab:benchmark_times}. In each sum, the first value corresponds to the simulation time, and the second to the STL verification time.

%
\begin{table}[h]
\centering
\caption{Computation time (in seconds) for two verification problems. The first was solved only with Dynibex, while the second compares Dynibex and CORA.}
\label{tab:benchmark_times}
\vspace{-1ex}
\begin{tabular}{|l|c|c|}
\hline
\multicolumn{3}{|c|}{$([\tilde{y}], t = 2.63s) \models \phi$ (Dynibex only)} \\
\hline
 & Uncertainty Tracking & Incremental Only \\
\hline
Dynibex & 0.485 + 0.222 = 0.707s & 3.37 + 0.15 = 3.52s \\
\hline
\end{tabular}

\vspace{0.3cm}

\begin{tabular}{|l|c|c|}
\hline
\multicolumn{3}{|c|}{$([\tilde{y}], t = 3s) \models \phi$ (Dynibex vs CORA)} \\
\hline
 & Uncertainty Tracking & Incremental Only \\
\hline
Dynibex & 0.600 + 0.156 = 0.756s & 4.0 + 0.15 = 4.15s \\
CORA    & --- & 11.9 + 4.2 = 16.1s \\
\hline
\end{tabular}
\end{table}
\vspace{-1ex}

In this case, verifying the formula at $t=2.63\,\mathrm{s}$ is among the most precision-demanding points. To obtain a conclusive verification result, we increased the precision criterion for the LTE by a factor of $10^{-5}$ and employed a higher-order integration scheme. Consequently, the simulation took approximately $7\times$ longer than with a less precise configuration. The increase in verification time (from $0.15 \mathrm{s}$ to $0.222 \mathrm{s}$) under uncertainty tracking is due to repeated bisection and contraction of the reachable sets during the STL evaluation.

From a broader perspective, at a less demanding time point ($t=3 \,$s), we compared execution times between our tracking implementation and CORA, using identical initial time steps (Table~\ref{tab:benchmark_times}). For each approach, we reported the fastest time required to reach a conclusive result, our tracking method is about $21\times$ faster than the incremental one using CORA. 
This performance gain may enable precise, on-the-fly verification of nonlinear behaviors against STL specifications in online monitoring contexts.

%
\section{Discussion}
One limitation of our approach lies in the propagation of uncertainty through unitary signals under the disjunction ($\lor$) operator. Contributing reachable set indices are merged across the entire resulting unitary signal, potentially leading to over-approximation. As shown in Fig.~\ref{tracking logic}, indices (1,2) affect only part of the signal but are attributed to the whole, which may trigger unnecessary refinements in early iterations, especially for complex formulas. 
However, this limitation is inherent to the propagation process within unitary decomposition.
As future work, incorporating time-dependence shifting into the markers could help reduce this over-approximation.
%

\section{Conclusion}
%
This work introduces a novel method for managing uncertainty in STL verification by embedding Boolean interval arithmetic within a reachability analysis framework. Our method systematically tracks uncertainty within the satisfaction tree, distinguishing between uncertainties arising from incomplete observations and those caused by reachable set over-approximation. This enables an offline/online refinement strategy, reducing unnecessary computations while ensuring precise verification.

A key feature of our method is its ability to contract and resample only the necessary reachable sets, thereby refining STL satisfaction evaluation. Through an experimental case study, we demonstrate the effectiveness of our approach in reducing uncertainty.

\section*{ACKNOWLEDGMENT}
The authors acknowledge support from the CIEDS\footnote{CIEDS: French Interdisciplinary Center for Defense and Security.} with the STARTS project.


{\tiny
\bibliographystyle{IEEEtran}
\bibliography{refs.bib}
}

\end{document}